\documentclass[aps,prl,twocolumn,amsmath,amssymb,showpacs,floatfix]{revtex4}

\usepackage[dvipsnames,rgb,dvips]{xcolor}
\usepackage{overpic}

\usepackage{graphicx,mathrsfs,psfrag}
\usepackage{dcolumn}

\newcommand{\obs}[1]{{#1}}

\renewcommand{\cite}{\citep}
\renewcommand{\cite}{\citep}
\newcommand{\ve}[1]{\ensuremath{\mbox{\boldmath$#1$}}}
\newcommand{\ma}[1]{\ensuremath{\mathbb{#1}}}
\newcommand{\T}{^{\rm T}}

\newcommand{\dl}{\ensuremath{d_{\rm L}}}
\newcommand{\eqnlab}[1]{\label{eq:#1}}

\newcommand{\Eqnref}[1]{Eq.~(\ref{eq:#1})}
\newcommand{\Figref}[1]{Fig.~\ref{fig:#1}}

\newcommand{\marker}[1]{\protect\includegraphics[width=2mm,clip]{mark#1.pdf}}

\newcommand{\gghat}{\ensuremath{\hat{\ve g}}}

\DeclareMathOperator{\tr}{Tr}
\DeclareMathOperator{\ku}{Ku}

\DeclareMathOperator{\G}{\Phi}

\begin{document}
\title{Preferential sampling and small-scale clustering of gyrotactic microswimmers in turbulence\footnote{Version accepted for publication (postprint) in Phys. Rev. Lett. {\bf 116}, 108104 (2016)}}
\author{K. Gustavsson$^{1,2)}$, F. Berglund$^{1)}$, P.R. Jonsson$^{3)}$,
and B. Mehlig$^{1)}$ }
\affiliation{\mbox{}$^{1)}$Department of Physics, Gothenburg University, SE-41296 Gothenburg, Sweden}
\affiliation{\mbox{}$^{2)}$Department of Physics and INFN, University of Rome 'Tor Vergata', 00133 Rome, Italy}
\affiliation{\mbox{}$^{3)}$Department of Biological and Environmental Sciences - Tj\"arn\"o, SE-45296 Str\"omstad,Sweden}

\begin{abstract}
Recent studies show that spherical motile micro-organisms in turbulence subject to gravitational torques gather in down-welling regions of the turbulent flow. By analysing a statistical model we analytically compute how shape affects the dynamics, preferential sampling, and  small-scale spatial clustering. We find that oblong organisms may spend more time in up-welling regions of the flow, and that all organisms are biased to regions of positive fluid-velocity gradients in the upward direction. We analyse small-scale spatial clustering  and find that oblong particles may either cluster more or less than spherical ones, depending on the strength of the gravitational torques.  
\end{abstract}

\maketitle

Patchiness in suspensions of micro-organisms
is frequently observed on a range of spatial scales.
The underlying mechanisms differ, depending on the properties
of the micro-organisms, and upon the spatial scale. Patchiness
can be caused by
density stratification and vertical shears \cite{FJ2008}, by predator-prey cycles, or by interactions between the organisms and water-column gradients -- in light, chemistry, turbulence,  and in hydrostatic pressure~\cite{Fol99}.  Patchiness is important
because many biological processes (mating, feeding, predation)
rely on individual encounters \cite{Kor08}, and the encounter rate is strongly influenced
by small-scale number-density fluctuations.

Gravitaxis may cause such inhomogeneities in the spatial distribution  of motile micro-organisms.  Density- or drag-asymmetries of the body give rise to torques
affecting the swimming direction \cite{Roberts1970,Jonsson1989,RD2002}.
When the effects of gyrotactic torques and
fluid-velocity gradients balance,  inhomogeneities may form in the spatial distribution, as shown by the micro-alga {\em Chlamydomonas nivalis} swimming up against a down-welling pipe flow. The micro-algae  gather in the centre of the pipe where the down-welling velocity is largest \cite{Kes85}.
Gyrotaxis may trap motile organisms in macroscopic shear gradients
\cite{Dur09,San14}, and fluctuating vorticity may cause patchiness \cite{Mitchell1990}.
This is confirmed by recent direct numerical simulations (DNS) of motile, spherical micro-organisms in turbulence \cite{Dur13} revealing that the organisms are more likely to be found in down-welling regions of the turbulent flow, they \lq preferentially sample\rq{} such regions.

These results raise three fundamental questions that we address and answer in this Letter.
First, how does shape affect the dynamics in turbulence of motile micro-organisms subject to gyrotaxis?
In Ref.~\cite{Dur13} the organisms were assumed to be spherical. Non-spherical
organisms respond not only to turbulent vorticity but also to
turbulent strain \cite{Jef22,Par12,Gus14,Byr14}.
This causes passive rods to exhibit intricate
orientational patterns on the surface of turbulent and other complex
flows \cite{Wil09,Wil10a,Wil11}.
Also, shape strongly affects the trajectories of active particles in model flows \cite{Tor07,Dur11,Khu12}, and recent
DNS indicate that prolate gyrotactic organisms cluster less than spherical
ones  when gyrotaxis is strong \cite{Zha14}.
Second, where do the organisms go in turbulence? Are there circumstances where the organisms may not gather in down-welling regions, or where other mechanisms of preferential sampling may apply?
Third, the fact that orgamisms tend to gather in certain regions of the flow (preferential sampling) does not explain which mechanisms actually cause them to get in contact.
To determine these one must follow the dynamics of two organisms that are initially very close together, and determine whether they tend to approach further or move apart. We refer to the resulting small-scale spatial fluctuations in the number density as \lq small-scale clustering\rq{}.

{\em Statistical model}.
To answer these questions we use a simplified model~\cite{Kes85,Dur13,Zha14} for  the translation and rotation
of small axisymmetric active particles subject to turbulence and gyrotaxis:
\begin{equation}
\label{eq:dotr}
\dot{\ve r} \equiv \ve v =
\ve u(\ve r,t) + v_{\rm s} {\ve n} \quad \mbox{and}\quad\dot {\ve n} = \ve \omega(\ve r,t) \wedge \ve n\,.
\end{equation}
Dots denote derivatives w.r.t. time $t$,
$\ve r$ is the particle position, and $\ve u$ is the flow velocity. Each particle
swims with constant speed $v_{\rm s}$ in the direction $\ve n$ of its symmetry axis ($|\ve n|=1$).
The angular velocity of the particle is
\begin{equation}
\label{eq:dotn}
\ve \omega(\ve r,t) \!=\!  (\hat{\ve g}\wedge \ve n)/(2\mathscr B)\!+\!\ve \Omega(\ve r, t) \!+ \!\Lambda \ve n \wedge  \big[\ma S(\ve r, t)\ve n\big]\,.
\end{equation}
The first term on the r.h.s describes gyrotaxis.
The unit vector $\hat{\ve g}$ points in the direction $-{\ve e}_z$ of gravity, and
$\mathscr{B}$ is the reorientation time \cite{Kes85,Dur13}. It depends on the mass distribution within the particle,
and on its shape through hydrodynamic resistance.
The other terms  on the r.h.s. of Eq.~(\ref{eq:dotn}) represent the effect of the turbulent velocity gradients upon
the particle orientation \cite{Jef22}: $\ve \Omega=(\ve \nabla \wedge \ve u)/2$, and
 $\ma S$ is the symmetric part of the matrix $\ma A$ of fluid-velocity gradients.
The parameter  $\Lambda$ characterises particle shape: $\Lambda\!=\!0$ for spheres, and $\Lambda\!=\!1$  for infinitely thin rods.
Eq.~(\ref{eq:dotn}) disregards  turbulent accelerations. In most marine conditions this is an excellent approximation \cite{Del14}.
We model the dissipative range of turbulence by incompressible, homogeneous, isotropic Gaussian random
functions with typical length- , time- , and speed-scales $\eta$, $\tau$, $u_0$~\cite{Gus14f}.
This neglects inertial-range properties which may become important for
particles that are larger than the Kolmogorov length \cite{Pec12}.
We note that the dissipative-range turbulent fluctuations are universal \cite{Sch14}, but they are not Gaussian. We comment on this difference between turbulence and the statistical model below.

There are four dimensionless parameters:
the shape parameter $\Lambda$, the reorientation time $\Psi = \mathscr{B}/\tau$,
the swimming speed $\G = v_{\rm s}\tau/\eta$, and the Kubo number $\ku = u_0 \tau/\eta$.
We vary the parameters independently,
keeping $\mathscr{B}$ constant as $\Lambda$ is changed.
$\ku$ is a dimensionless measure of the correlation time of the flow.

Our choice of the dimensionless parameters is dictated by the method (explained below). DNS employ different de-dimensionalisations~\cite{Dur13}:
${\mathscr B}$ by the Kolmogorov time $\tau_{\rm K}\equiv 1/\sqrt{\tr\langle\ma A\ma A\T\rangle_\infty}$, and $v_{\rm s}$ by the corresponding Kolmogorov speed $u_{\rm K}$.
 Our dimensionless parameters translate to those used in the DNS as $\Psi_{\rm DNS}\sim\ku\Psi$ and $\Phi_{\rm DNS}\sim  \G/\ku$.
We expect that the statistical-model results become independent of $\ku$ at large $\ku$ and qualitatively agree with DNS results \cite{Gus14f}.

Typical values of $\mathscr{B}$, $v_{\rm s}$
are given in Ref.~\cite{Jonsson1989}:
${\mathscr B}\sim 1$--$5\,$s and $v_{\rm s}\sim 0.1$--$1\,$mm/s.
Typical ocean dissipation rates are
$\varepsilon\sim 1$--$10^{2}\,$mm$^2$/s$^3$ for surface water \cite{Yam96},
giving Kolmogorov times, lengths, and speeds in the range $\tau_{\rm K}\sim 0.1$--$1\,$s,
$\eta\sim 0.3$--$1\,$mm, and $u_{\rm K}\sim$ $1$--$3\,$mm/s.
These estimates yield $\Psi_{\rm DNS}\sim 1$--$50$ and $\Phi_{\rm DNS}\sim 0.03$--$1$.
In Ref.~\cite{Gar89} smaller dissipation rates, $\varepsilon\sim 10^{-4}\,$mm$^2$/s$^3$, are quoted
for the very deep sea. This extends the ranges to $\Psi_{\rm DNS}\sim 0.01$--$50$ and $\Phi_{\rm DNS}\sim 0.03$--$10$.

{\em Method.}
Eqs.~(\ref{eq:dotr},\ref{eq:dotn}) can be solved
by iteratively refining approximations for the path a particle takes through the flow~\cite{Gus11a,Gus14e,Gus14f}. This results in expansions of steady-state averages in powers of $\ku$ and allows to determine how the remaining parameters ($\G$, $\Psi$, and $\Lambda$) affect preferential sampling and small-scale clustering. The details of this calculation are given in the Supplemental Material \cite{supp}. Here we outline the essential steps.
First, to consistently track the orders in the expansion
we de-dimensionalise $t'\!=\!t/\tau, \ve r'\!=\! \ve r/\eta, \ve u'\!= \!\ve u/u_0$.
Second, we expand the dynamics of the vector $\ve n$ in powers of $\ku$:
\begin{equation}
\ve n(t')=\sum_{q=0}^\infty\ve n_q(t')\ku^q\,.
\end{equation}
Inserting this ansatz into (\ref{eq:dotr},\ref{eq:dotn}) and identifying terms of order $\ku^q$
yields equations for ${\ve n}_q$ that can be solved in terms of $\ve n_p$ for  $p<q$.
The lowest-order solution in $\ku$ is just $\ve n_0=-\gghat$. This yields
a lowest-order deterministic approximation
for the particle position at time $t'$:
\begin{equation}
{\ve r}'_{\!\rm det}(t')=\ve r_0'-\G\gghat \,t'\,.
\end{equation}
Third, we expand Eqs.~(\ref{eq:dotr},\ref{eq:dotn})
in terms of deviations from the deterministic trajectory $\delta\ve r'({t'})\equiv\ve r'({t'})-{\ve r}'_{\!\rm det}(t')$
precisely as described in Ref.~\cite{Gus14f}. In the fourth and final step we average over the fluid-velocity fluctuations in the statistical model. In the remainder of this Letter we summarise the results obtained in this way.

\begin{figure}[t]
  \begin{overpic}[width=8.5cm,clip]{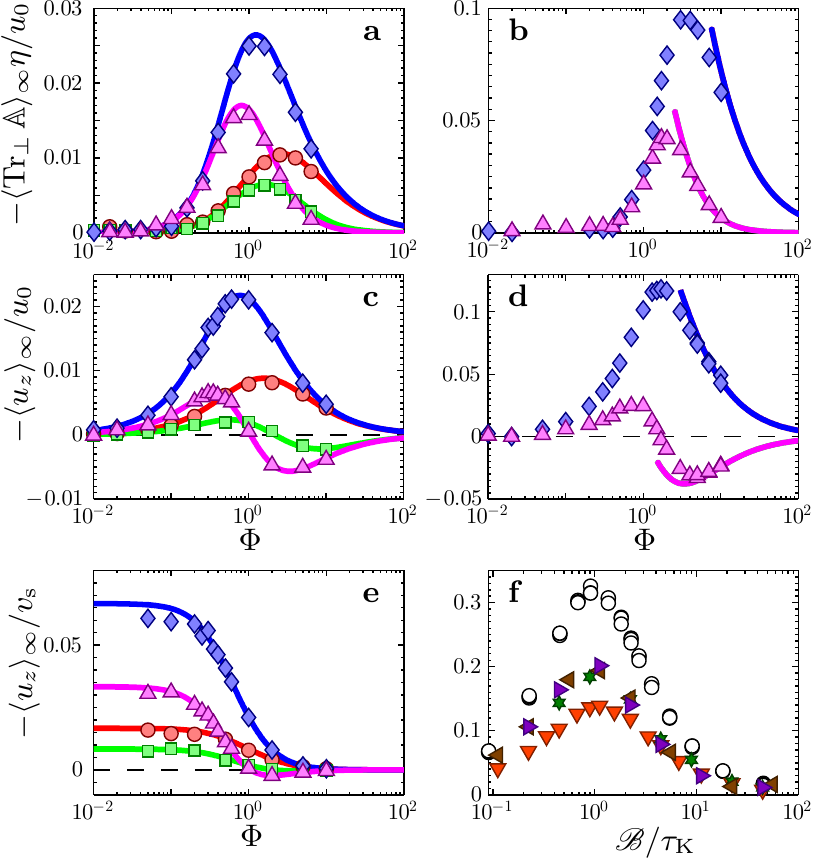}
  \end{overpic}\hspace*{2mm}
  \caption{\label{fig:preferential_sampling} Preferential sampling.
{\bf a} $\langle \tr_\perp \ma A\rangle_\infty$
as sampled by the particles. Results of simulations of statistical model: symbols.
Eq.~(\ref{eq:S}): lines. Parameters: $\ku = 0.1$,
$\Psi = 0.1$, $\Lambda=0$ ($\marker{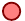}$), $\Psi=1$, $\Lambda =0$ ($\marker{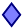}$),
 $\Psi= 0.1$, $\Lambda = 1$ ($\marker{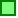}$) $\Psi=1,\Lambda=1$ ($\marker{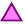}$).
{\bf b} The same but for $\ku=1$ (theory only for large $\G$). {\bf c} $\langle u_z\rangle_\infty$
as sampled by the particles for $\ku=0.1$. {\bf d} Same but for $\ku=1$.
{\bf e} $\langle u_z\rangle_\infty/v_{\rm s}$ as a function of $\Phi$ for $\ku=0.1$.
{\bf f} $\langle u_z\rangle_\infty/{v}_{\rm s}$ versus ${\mathscr{B}}/\tau_{\rm K}$ for $\Lambda=0$ and $\ku=1$ ($\marker{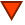}$), $\ku=2$ ($\marker{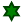}$), $\ku=5$ ($\marker{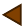}$), $\ku=10$ ($\marker{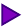}$).
Hollow markers show DNS data from Fig.~3{\bf d} in Ref.~\cite{Dur13} (at ${\rm Re}_\lambda=64$).
All data are for values of $\Phi$ from the small-$\Phi$ plateau observed in DNS \cite{Dur13}, and also in the statistical model (for $\ku=0.1$ this plateau is shown in panel {\bf e}, for $\ku=1$ in Fig.~S1 in the Supplemental Material).
Panels {\bf a},{\bf c},{\bf e} are for $d=2$, panels {\bf b},{\bf d},{\bf f} for $d=3$.
}
\end{figure}

{\em Preferential sampling.}
Consider the steady-state averages
of the $z$-component $u_z$ of the fluid velocity and of its gradient, $A_{zz}$, both evaluated at the particle position. Analytical results
for these averages are derived to lowest order in $\ku$ in the Supplemental Material~\cite{supp}, 
Eqs.~(S15) and (S16). These expressions are plotted in \Figref{preferential_sampling}. Here we quote only limiting results.
For small $\G$  we have
\begin{subequations}
\label{eq:ps}
\begin{align}
\label{eq:S}
\langle A_{zz}\rangle_\infty\frac{\eta}{u_0}&\!\sim\!\ku\G^2\frac{d(1\!-\!\Lambda)\!+\!2(\Lambda\!+\!2)}{d}\,\frac{\Psi(4\Psi+\!1)}{(2\Psi\!+\!1)^2}\,,\\
\label{eq:uz}
\frac{\langle u_z\rangle_\infty}{u_0}&\sim -\,\ku\G\frac{d(1-\Lambda)+2}{d}\,\frac{\Psi}{2\Psi+1}\,,
\end{align}
\end{subequations}
$d$ is the spatial dimension.
For large $\G$ we find
\begin{subequations}
\label{eq:ps_largeG}
\begin{align}
\label{eq:S_largeG}
\langle A_{zz}\rangle_\infty\frac{\eta}{u_0}&\sim \frac{\ku}{\G}\,\frac{d+1}{2d}(1-\Lambda)\sqrt{\frac{\pi}{2}}\,,\\
\label{eq:uz_largeG}
\frac{\langle u_z\rangle_\infty}{u_0}&\sim \frac{\ku}{\G}\,\frac{(d(\Lambda-1) + 2\Lambda)}{2d}\,.
\end{align}
\end{subequations}
What can we learn from these analytical results?
Eqs.~(\ref{eq:S}) and (\ref{eq:S_largeG}) show that the particles collect in the sinks of the transversal flow-velocity field,
$\tr_\perp \ma A  \equiv -A_{zz} < 0$.
This is because gyrotaxis breaks up-down symmetry: when $\Psi$ is small the particles swim essentially upwards (in the ${\bf e}_z$-direction), and
gather in transversal sinks irrespective of their shape and swimming speed.
Simulations (Fig.~\ref{fig:preferential_sampling}{\bf a}) confirm the theory.

Motivated by Kessler's study in pipe flows \cite{Kes85} the authors of Ref.~\cite{Dur13}
concluded that spherical particles preferentially sample down-welling regions also in turbulence.
This is not in contradiction with the result discussed above because particles may preferentially sample different observables. In
fact Eqs.~(\ref{eq:uz}) and (\ref{eq:uz_largeG}) explain that spherical particles are biased towards down-welling regions (as observed in DNS \cite{Dur13}),
in addition to sinks in the transversal flow. But (\ref{eq:uz_largeG}) also shows that elongated particles [$\Lambda>d/(d+2)$] preferentially sample up-welling regions for large enough $\G$. This is seen in Fig.~\ref{fig:preferential_sampling}{\bf c} which shows $\langle u_z\rangle_\infty$ (Eq.~(S15) in the Supplemental Material). for $\ku=0.1$ and $d=2$ as a function of~$\G$.
Also shown are results of statistical-model simulations, in excellent agreement with theory.
Fig.~\ref{fig:preferential_sampling}{\bf d} shows that the same conclusions hold in three spatial dimensions for $\ku=1$.
Rods sample upwelling regions when $\G$ is larger than (approximately) unity.

It is remarkable that the shapes of the curves at $\ku=0.1$ are very similar to those at large $\ku$. This means that the small-$\ku$ theory
 {\em qualitatively} explains what is observed in the statistical-model simulations at large $\ku$ and in DNS for spherical particles \cite{Dur13}.

In the limit of large $\Phi$ particles swim rapidly upwards and experience the flow as a white-noise signal (as in rapid gravitational settling \cite{Gus14e}).  This limit is universal, particles in {\em any} homogeneous, isotropic and incompressible flow show preferential sampling according to \Eqnref{ps_largeG}.
This means that the small-$\ku$ theory should describe results of statistical-model simulations at $\ku=1$ quantitatively for large $\G$.  This is confirmed by Figs.~\ref{fig:preferential_sampling}{\bf b},{\bf d}.

For small $\Phi$ DNS \cite{Dur13} show that the average of $u_z$ is proportional to $\G$ for small $\G$, so that $\langle u_z\rangle_\infty/\G$ is constant. Eq.~(\ref{eq:uz}) shows this behaviour, in good agreement with simulations (Fig.~\ref{fig:preferential_sampling}{\bf e}).

We conclude with a quantitative comparison of  statistical-model and DNS results \cite{Dur13}. As an example consider  the dependence of $\langle u_z\rangle_\infty$  on the reorientation time  $\mathscr{B}$.
Fig.~\ref{fig:preferential_sampling}{\bf f} shows that the statistical-model result becomes independent of $\ku$ for large $\ku$,
and that it reproduces the DNS results fairly well, it explains the $\mathscr{B}/\tau_K$-dependence of 
$\langle u_z\rangle_\infty$ of the DNS results up to a prefactor of order unity.
This factor is due to the fact that fully-developed turbulent velocity fluctuations in the dissipative range differ from those in the statistical model: they are not Gaussian, more persistent, and the probability of straining regions to occur is higher \cite{Gus14f}. 

{\em Small-scale clustering.} Which mechanisms cause two particles caught
in the same flow region to actually collide?
This is a two-particle problem, only indirectly related to preferential sampling. Fluctuations in the separations between nearby particles are determined by
the dynamics of the particle-velocity gradients $\partial v_i/\partial r_j$.
Small-scale clustering 
occurs where $\ve \nabla \cdot \ve v<0$.
We have computed $\langle\ve \nabla \cdot \ve v\rangle_\infty$ to lowest order in $\ku$. The 
result is quite lengthy [Eq.~(S32) in the Supplemental Material]. For small $\Phi$ the full expression 
simplifies to:
\begin{align}
\label{eq:divv_smallPhi}
\langle \ve \nabla \cdot \ve v\rangle_\infty\eta/u_0&\sim-\ku\, (\Phi\Psi)^2 B_d(\Lambda)\quad\mbox{for}\quad\Phi \ll 1\,,
\end{align}
with $B_d(\Lambda)\equiv [(d\!+\!2)(d\!+\!4) \!-\! 2d(d\!+\!4)\Lambda+ (4\!+\!2d\!+d^2)\Lambda^2\big]/d$.  Since $B_d(\Lambda)>0$, Eq.~(\ref{eq:divv_smallPhi}) implies small-scale clustering. 
For spherical particles ($\Lambda=0$) the quadratic dependence of 
$\langle \ve \nabla \cdot \ve v\rangle_\infty$ on $\Phi\Psi$ was derived in Ref.~\cite{Dur13} 
(and also in Ref.~\cite{Fou15}): expanding  Eqs. (\ref{eq:dotr},\ref{eq:dotn}) for $\mathscr{B}\ll\tau$  gives
\begin{align}
\ve \nabla \cdot \ve v&\sim v_{\rm s} \mathscr{B}\big[{-(1\!+\!\Lambda)}\,\partial_{z}^2 u_{z}+{(1\!-\!\Lambda)}(\partial_{z}^2 u_{z}\!-\!\Delta u_{z})\big]\,.
\eqnlab{TrZSmallPsi}
\end{align}
Substituting $\Lambda=0$ yields Eq.~(6) of Ref.~\cite{Dur13}, and averaging Eq.~(\ref{eq:TrZSmallPsi}) along particle paths results in Eq.~(\ref{eq:divv_smallPhi}).
The factor $v_{\rm s}\mathscr{B}$ in (\ref{eq:TrZSmallPsi}) corresponds to one factor of $\G\Psi$ in~(\ref{eq:divv_smallPhi}). The second
factor of $\G\Psi$ comes from averaging the velocity derivatives in Eq.~(\ref{eq:TrZSmallPsi}).
We note that $\tr_\perp \ma A$ does not figure in Eq. (\ref{eq:TrZSmallPsi}):
preferential sampling of sinks
in the flow-velocity field perpendicular to gravity does not contribute
to small-scale clustering, showing that the two effects are distinct \cite{Gus14f}.

Expanding the full result (S32) for large $\Phi$ gives:
\begin{align}
 \langle \ve \nabla \cdot \ve v\rangle_\infty\eta/u_0&\sim-\ku\, \Phi\Psi^2 \,E_d(\Lambda)
\quad\mbox{for}\quad\Phi\gg 1\,,
 \label{eq:divv_largePhi}
\end{align}
with $E_d(\Lambda)\equiv\sqrt{\pi/2}(d+1)(d+3)(\Lambda-1)^2/d$. For spherical particles the $\Phi\Psi^2$-dependence was derived in Ref.~\cite{Fou15}.

Let us now analyse the shape dependence of Eq.~(\ref{eq:divv_smallPhi}). The $\Lambda$-dependence
of $B_d(\Lambda)$ explains that rods ($\Lambda=1$) cluster less than spheres ($\Lambda=0)$, consistent with the DNS results reported in Ref. \cite{Zha14}. 
But when gyrotaxis is weak, spheres
 are essentially randomly oriented, unlike neighbouring rods that are aligned by turbulent shears.
In this limit motile  rods must cluster more than spheres.
This  is demonstrated below, but it is
not captured by Eqs.~(\ref{eq:divv_smallPhi}) and (\ref{eq:divv_largePhi}) which must fail for large $\Psi$ because the limit $\Psi\to\infty$ is singular, and due to the occurrence of singularities in the dynamics of 
the gradients of $\ve n$ at large but finite values of $\Psi$ (Supplemental Material).
The first caveat also applies to Eqs.~(\ref{eq:ps}) and (6).

\begin{figure}[t!]
\begin{overpic}[width=8.7cm,clip]{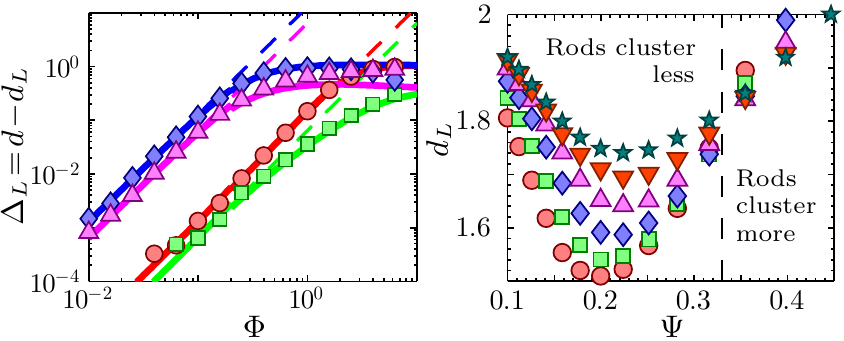}
  \put(12,9){{{\bf a}}} \put(62,9){{{\bf b}}}
  \end{overpic}\hspace*{2mm}
\caption{\label{fig:dL}
{\bf a} Fractal dimension deficit
 $\Delta_L \equiv d-\dl$
 for $d=2$, $\ku=0.1$. Numerical simulations of statistical model for $\Psi=0.1,\Lambda=0$ ($\marker{1}$), $\Psi = 1,\Lambda=0$ ($\marker{3}$),
 $\Psi =0.1,\Lambda=1$ ($\marker{2}$), and $\Psi=1,\Lambda=1$ ($\marker{4}$).
Theory [Eqs.~(\ref{eq:dl}), (S31), and (S32)] including the $\ku^4$-contribution to Eq.~(S31): solid lines.
Asymptote $\propto (\G\Psi)^2$: dashed lines.
{\bf b} Numerical simulations,  $\dl$
 for $d=2$, $\G=1$, $\ku=1$, $\Lambda=0$ ($\marker{1}$), $0.2$ ($\marker{2}$), $0.4$ ($\marker{3}$), $0.6$ ($\marker{4}$),
$0.8$ ($\marker{5}$), $1$ ($\marker{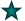}$).
}
\end{figure}

{\em Fractal dimension.} DNS show \cite{Dur13} that the small-scale
spatial patterns of motile gyrotactic organisms
are fractal. This may substantially enhance their encounter rates \cite{And07}.
We analyse the fractal patterns for finite but small $\ku$,
 in two dimensions. We expect qualitatively the same result in three dimensions. The fractal patterns are characterised by \lq Lyapunov exponents\rq{}
$\lambda_1$ and $\lambda_2$
\begin{align}
\lambda_1&\!\equiv\!\lim_{t\to\infty}\!t^{-1}\ln \frac{\scriptstyle R({t})}{\scriptstyle R(0)}
\!\!\!\quad\mbox{and}\quad \!\!\!\lambda_1\!+\!\lambda_2\!\equiv\!\lim_{t\to\infty}\!
t^{-1}\ln \frac{\scriptstyle {\cal A}({t})}{\scriptstyle {\cal A}(0)}\,.
\end{align}
These exponents quantify the expansion (contraction) rates of
the distance $R({t})$ between two initially nearby
particles, and of the area element ${\cal A}({t})$
spanned by the separation vectors between three nearby particles. The fractal Lyapunov dimension
is defined by~\cite{Kap79,Gus14f}
\begin{equation}
\label{eq:dl}
\dl\equiv 1-\lambda_1/\lambda_2\,,
\end{equation}
assuming  $\lambda_1>0$ and $\lambda_1+\lambda_2<0$. When $\dl < 2$ fractal clustering occurs.
To evaluate $d_{\rm L}$ we use $\lambda_1+\lambda_2
=\langle \ve \nabla \cdot \ve v\rangle_\infty$\obs{, Eq.~(S32),} and compute $\lambda_1$ to order $\ku^4$. The result is lengthy, in the Supplemental Material~\cite{supp} we quote the result to order $\ku^2$, Eq.~(S31). To this order it is independent of $\Psi$ and $\Lambda$. For small values of $\G$ Eq.~(S31) simplifies to $\lambda_1\tau\sim \ku^2(1 - 3\G^2)$ for $d=2$. Together with (\ref{eq:divv_smallPhi}) this implies
$\Delta_{\rm L} \equiv d-\dl \sim \Phi^2\Psi^2$ consistent with the results of Refs.~\cite{Dur13,Fou15} for spherical particles.
Fig.~\ref{fig:dL}{\bf a} shows the analytical result for  $\dl$
as a function of $\G$. It is in good agreement with numerical simulations of the statistical model ($d=2$) for $\ku=0.1$. We see that spherical organisms cluster more than rods. As explained above this is expected  for strong gyrotaxis.

But when the effect of the gravitational torque is small
then prolate organisms cluster more: in the absence of gyrotaxis, rotational symmetry ensures that active spherical particles remain uniformly distributed, but rod-like particles show fractal clustering. Panel {\bf b} in Fig. \ref{fig:dL} demonstrates this cross-over. It shows $\dl$  for $\G=1$, $\ku=1$ as a function of $\Psi$.
We arrive at qualitatively similar conclusions by numerically computing the fractal correlation dimension $d_2$. But  the numerical values found for $d_2$ differ from $\dl$
This  shows that the spatial distribution is multifractal \cite{Gus14f}.

{\em Conclusions.}
First, our statistical-model calculations  explain how
the dynamics of gyrotactic motile micro-organisms depends
on the dimensionless parameters of the problem: $\Lambda$ (shape), $\G$ (swimming speed), and $\Psi$
(reorientation time).
Second, we find that the particles tend to preferentially sample positive values of $A_{zz}$, corresponding to sinks in the transversal flow, regardless of shape. We predict that this must also be observed in DNS, it is simply a consequence of the fact that gravity breaks the symmetry of the problem. At the same time our calculations show that spherical particles are more often
found in regions where $u_z$ is negative, explaining the \obs{behaviour} found in DNS \cite{Dur13}. But
our calculations also predict that rod-like particles preferentially sample up-welling regions of homogeneous isotropic flows such as turbulence, provided that they swim fast enough.
Third, we have analytically computed how the degree of small-scale spatial clustering
depends on particle shape. This is important because small-scale fractal clustering may enhance particle-encounter rates. We find a transition that we predict to be observable in DNS as well:
when gyrotaxis is strong (small $\Psi$) oblong particles cluster less than spherical ones, while
at large $\Psi$ the opposite is true.

Our calculations also show that singularities in the motion of nearby micro-organisms occur, 
much like  \lq caustics\rq{} for heavy particles in turbulence  \cite{Wil05,Fal02,Gus12,Meh04}.
We predict that such singularities  must also be observed in the DNS of \obs{gyrotactic microswimmers in turbulence.}
\obs{It is of interest to estimate how often the singularities occur
because their effect may modify the predictions of phenomenological models for encounter rates~\cite{Rot88}.}

The analytical results obtained in this Letter were derived for  small $\ku$ (or large  $\G$).
But we have shown  that our analytical results and  the corresponding mechanisms \obs{qualitatively}
explain what is observed in DNS, and explain also the results of statistical-model
simulations at \obs{large values of $\ku$.}
We find \obs{fairly} good \obs{quantitative} agreement between our statistical-model calculations and DNS results for fully
developed turbulence. To achieve even better quantitative agreement with the DNS would 
require to account for \obs{the universal non-Gaussian small-scale fluctuations 
of fully developed turbulence}~\cite{Sch14}.

But the fluctuations of the unsteady ocean are neither fully-developed turbulent, nor are they Gaussian.
Therefore the fact that the much simpler Gaussian statistical model explains the dynamics observed in DNS of fully developed turbulence \cite{Dur13} shows that the analytical theory (and the underlying mechanisms) describe robust behaviour, \obs{that} must be taken into account in the analysis of patchiness and encounter rates of motile micro-organisms in the ocean.

{\em Acknowledgments.}
We thank the authors of Ref.~\cite{Dur13} for permission to reproduce their data in Fig.~3{\bf d}.
This work was supported by Vetenskapsr\aa{}det (VR),
 by a Linnaeus-grant from VR and
  Formas, by
  the G\"oran Gustafsson Foundation for Research in
 Natural Sciences and Medicine, and 
by the grant {\em Bottlenecks for particle growth in turbulent aerosols} from the Knut and Alice Wallenberg Foundation, Dnr. KAW 2014.0048.
K.G. acknowledges partial funding from the European Research Council under the European Community’s Seventh Framework Programme, ERC Grant Agreement N. 339032.
  The numerical computations used resources
 provided by the Centre for Scientific and Technical Computing at Chalmers University of Technology in Gothenburg 
(Sweden) and the Swedish National Infrastructure for Computing.


\begin{thebibliography}{36}
\expandafter\ifx\csname natexlab\endcsname\relax\def\natexlab#1{#1}\fi
\expandafter\ifx\csname bibnamefont\endcsname\relax
  \def\bibnamefont#1{#1}\fi
\expandafter\ifx\csname bibfnamefont\endcsname\relax
  \def\bibfnamefont#1{#1}\fi
\expandafter\ifx\csname citenamefont\endcsname\relax
  \def\citenamefont#1{#1}\fi
\expandafter\ifx\csname url\endcsname\relax
  \def\url#1{\texttt{#1}}\fi
\expandafter\ifx\csname urlprefix\endcsname\relax\def\urlprefix{URL }\fi
\providecommand{\bibinfo}[2]{#2}
\providecommand{\eprint}[2][]{\url{#2}}

\bibitem[{\citenamefont{Franks and Jaffe}(2008)}]{FJ2008}
\bibinfo{author}{\bibfnamefont{P.~J.~S.} \bibnamefont{Franks}}
  \bibnamefont{and} \bibinfo{author}{\bibfnamefont{J.~S.} \bibnamefont{Jaffe}},
  \bibinfo{journal}{J. Marine Sys.} \textbf{\bibinfo{volume}{69}},
  \bibinfo{pages}{254} (\bibinfo{year}{2008}).

\bibitem[{\citenamefont{Folt and Burns}(1999)}]{Fol99}
\bibinfo{author}{\bibfnamefont{C.~L.} \bibnamefont{Folt}} \bibnamefont{and}
  \bibinfo{author}{\bibfnamefont{C.~W.} \bibnamefont{Burns}},
  \bibinfo{journal}{Trends Ecol. Evol.} \textbf{\bibinfo{volume}{300}},
  \bibinfo{pages}{14} (\bibinfo{year}{1999}).

\bibitem[{\citenamefont{Ki\o{}rboe}(2008)}]{Kor08}
\bibinfo{author}{\bibfnamefont{T.}~\bibnamefont{Ki\o{}rboe}},
  \emph{\bibinfo{title}{A mechanistic approach to plankton ecology}}
  (\bibinfo{publisher}{Princeton University Press}, \bibinfo{address}{New
  Jersey, USA}, \bibinfo{year}{2008}).

\bibitem[{\citenamefont{Roberts}(1970)}]{Roberts1970}
\bibinfo{author}{\bibfnamefont{A.~M.} \bibnamefont{Roberts}},
  \bibinfo{journal}{J. Exp. Biology} \textbf{\bibinfo{volume}{53}},
  \bibinfo{pages}{687} (\bibinfo{year}{1970}).

\bibitem[{\citenamefont{Jonsson}(1989)}]{Jonsson1989}
\bibinfo{author}{\bibfnamefont{\obs{P.~R.}}~\bibnamefont{Jonsson}},
  \bibinfo{journal}{\obs{Mar.} Ecol. Prog. Ser.} \textbf{\bibinfo{volume}{52}},
  \bibinfo{pages}{39} (\bibinfo{year}{1989}).

\bibitem[{\citenamefont{Roberts and Deacon}(2002)}]{RD2002}
\bibinfo{author}{\bibfnamefont{A.~M.} \bibnamefont{Roberts}} \bibnamefont{and}
  \bibinfo{author}{\bibfnamefont{F.~M.} \bibnamefont{Deacon}},
  \bibinfo{journal}{J. Fluid. Mech.} \textbf{\bibinfo{volume}{452}},
  \bibinfo{pages}{405} (\bibinfo{year}{2002}).

\bibitem[{\citenamefont{Kessler}(1985)}]{Kes85}
\bibinfo{author}{\bibfnamefont{J.~O.} \bibnamefont{Kessler}},
  \bibinfo{journal}{Nature} \textbf{\bibinfo{volume}{313}},
  \bibinfo{pages}{218} (\bibinfo{year}{1985}).

\bibitem[{\citenamefont{Durham et~al.}(2009)\citenamefont{Durham, Kessler, and
  Stocker}}]{Dur09}
\bibinfo{author}{\bibfnamefont{W.~M.} \bibnamefont{Durham}},
  \bibinfo{author}{\bibfnamefont{J.~O.} \bibnamefont{Kessler}},
  \bibnamefont{and} \bibinfo{author}{\bibfnamefont{R.}~\bibnamefont{Stocker}},
  \bibinfo{journal}{Science} \textbf{\bibinfo{volume}{323}},
  \bibinfo{pages}{1067} (\bibinfo{year}{2009}).

\bibitem[{\citenamefont{Santamaria et~al.}(2014)\citenamefont{Santamaria,
  De~Lillo.~F., Cencini, and Boffetta}}]{San14}
\bibinfo{author}{\bibfnamefont{F.}~\bibnamefont{Santamaria}},
  \bibinfo{author}{\bibnamefont{De~Lillo.~F.}},
  \bibinfo{author}{\bibfnamefont{M.}~\bibnamefont{Cencini}}, \bibnamefont{and}
  \bibinfo{author}{\bibnamefont{Boffetta}}, \bibinfo{journal}{Phys. Fluids}
  \textbf{\bibinfo{volume}{26}}, \bibinfo{pages}{111901}
  (\bibinfo{year}{2014}).

\bibitem[{\citenamefont{Mitchell et~al.}(1990)\citenamefont{Mitchell, Okubo,
  and Fuhrmann}}]{Mitchell1990}
\bibinfo{author}{\bibfnamefont{J.~G.} \bibnamefont{Mitchell}},
  \bibinfo{author}{\bibfnamefont{A.}~\bibnamefont{Okubo}}, \bibnamefont{and}
  \bibinfo{author}{\bibfnamefont{J.~A.} \bibnamefont{Fuhrmann}},
  \bibinfo{journal}{Limnology and Oceanography} \textbf{\bibinfo{volume}{35}},
  \bibinfo{pages}{123} (\bibinfo{year}{1990}).

\bibitem[{\citenamefont{Durham et~al.}(2013)\citenamefont{Durham, Climent,
  Barry, De~Lillo, Boffetta, Cencini, and Stocker}}]{Dur13}
\bibinfo{author}{\bibfnamefont{W.~M.} \bibnamefont{Durham}},
  \bibinfo{author}{\bibfnamefont{E.}~\bibnamefont{Climent}},
  \bibinfo{author}{\bibfnamefont{M.}~\bibnamefont{Barry}},
  \bibinfo{author}{\bibfnamefont{F.}~\bibnamefont{De~Lillo}},
  \bibinfo{author}{\bibfnamefont{G.}~\bibnamefont{Boffetta}},
  \bibinfo{author}{\bibfnamefont{M.}~\bibnamefont{Cencini}}, \bibnamefont{and}
  \bibinfo{author}{\bibfnamefont{R.}~\bibnamefont{Stocker}},
  \bibinfo{journal}{Nature Comm.} \textbf{\bibinfo{volume}{4}},
  \bibinfo{pages}{2148} (\bibinfo{year}{2013}).

\bibitem[{\citenamefont{Jeffery}(1922)}]{Jef22}
\bibinfo{author}{\bibfnamefont{G.~B.} \bibnamefont{Jeffery}},
  \bibinfo{journal}{Proc. R. Soc. A} \textbf{\bibinfo{volume}{102}},
  \bibinfo{pages}{161} (\bibinfo{year}{1922}).

\bibitem[{\citenamefont{Parsa et~al.}(2012)\citenamefont{Parsa, Calzavarini,
  Toschi, and Voth}}]{Par12}
\bibinfo{author}{\bibfnamefont{S.}~\bibnamefont{Parsa}},
  \bibinfo{author}{\bibfnamefont{E.}~\bibnamefont{Calzavarini}},
  \bibinfo{author}{\bibfnamefont{F.}~\bibnamefont{Toschi}}, \bibnamefont{and}
  \bibinfo{author}{\bibfnamefont{G.~A.} \bibnamefont{Voth}},
  \bibinfo{journal}{Phys. Rev. Lett.} \textbf{\bibinfo{volume}{109}},
  \bibinfo{pages}{134501} (\bibinfo{year}{2012}).

\bibitem[{\citenamefont{Gustavsson
  et~al.}(2014{\natexlab{a}})\citenamefont{Gustavsson, Einarsson, and
  Mehlig}}]{Gus14}
\bibinfo{author}{\bibfnamefont{K.}~\bibnamefont{Gustavsson}},
  \bibinfo{author}{\bibfnamefont{J.}~\bibnamefont{Einarsson}},
  \bibnamefont{and} \bibinfo{author}{\bibfnamefont{B.}~\bibnamefont{Mehlig}},
  \bibinfo{journal}{Phys. Rev. Lett.} \textbf{\bibinfo{volume}{112}},
  \bibinfo{pages}{014501} (\bibinfo{year}{2014}{\natexlab{a}}).

\bibitem[{\citenamefont{Byron et~al.}(2015)\citenamefont{Byron, Einarsson,
  Gustavsson, Voth, Mehlig, and Variano}}]{Byr14}
\bibinfo{author}{\bibfnamefont{M.}~\bibnamefont{Byron}},
  \bibinfo{author}{\bibfnamefont{J.}~\bibnamefont{Einarsson}},
  \bibinfo{author}{\bibfnamefont{K.}~\bibnamefont{Gustavsson}},
  \bibinfo{author}{\bibfnamefont{G.~A.} \bibnamefont{Voth}},
  \bibinfo{author}{\bibfnamefont{B.}~\bibnamefont{Mehlig}}, \bibnamefont{and}
  \bibinfo{author}{\bibfnamefont{E.}~\bibnamefont{Variano}},
\bibinfo{journal}{Phys. Fluids} \textbf{\bibinfo{volume}{27}},
  \bibinfo{pages}{035101} (\bibinfo{year}{2015}).

\bibitem[{\citenamefont{Wilkinson et~al.}(2009)\citenamefont{Wilkinson,
  Bezuglyy, and Mehlig}}]{Wil09}
\bibinfo{author}{\bibfnamefont{M.}~\bibnamefont{Wilkinson}},
  \bibinfo{author}{\bibfnamefont{V.}~\bibnamefont{Bezuglyy}}, \bibnamefont{and}
  \bibinfo{author}{\bibfnamefont{B.}~\bibnamefont{Mehlig}},
  \bibinfo{journal}{Phys. Fluids} \textbf{\bibinfo{volume}{21}},
  \bibinfo{pages}{043304} (\bibinfo{year}{2009}).

\bibitem[{\citenamefont{Bezuglyy et~al.}(2010)\citenamefont{Bezuglyy, Mehlig,
  and Wilkinson}}]{Wil10a}
\bibinfo{author}{\bibfnamefont{V.}~\bibnamefont{Bezuglyy}},
  \bibinfo{author}{\bibfnamefont{B.}~\bibnamefont{Mehlig}}, \bibnamefont{and}
  \bibinfo{author}{\bibfnamefont{M.}~\bibnamefont{Wilkinson}},
  \bibinfo{journal}{Europhys. Lett.} \textbf{\bibinfo{volume}{89}},
  \bibinfo{pages}{34003} (\bibinfo{year}{2010}).

\bibitem[{\citenamefont{Wilkinson et~al.}(2011)\citenamefont{Wilkinson,
  Bezuglyy, and Mehlig}}]{Wil11}
\bibinfo{author}{\bibfnamefont{M.}~\bibnamefont{Wilkinson}},
  \bibinfo{author}{\bibfnamefont{V.}~\bibnamefont{Bezuglyy}}, \bibnamefont{and}
  \bibinfo{author}{\bibfnamefont{B.}~\bibnamefont{Mehlig}},
  \bibinfo{journal}{J. Fluid Mech.} \textbf{\bibinfo{volume}{667}},
  \bibinfo{pages}{158} (\bibinfo{year}{2011}).

\bibitem[{\citenamefont{Torney et~al.}(2007)\citenamefont{Torney, and
  Neufeld}}]{Tor07}
\bibinfo{author}{\bibfnamefont{C.} \bibnamefont{Torney}}, \bibnamefont{and}
  \bibinfo{author}{\bibfnamefont{Z.}~\bibnamefont{Neufeld}},
  \bibinfo{journal}{Phys. Rev. Lett.} \textbf{\bibinfo{volume}{99}},
  \bibinfo{pages}{78101} (\bibinfo{year}{2007}).

\bibitem[{\citenamefont{Durham et~al.}(2011)\citenamefont{Durham, Climent, and
  Stocker}}]{Dur11}
\bibinfo{author}{\bibfnamefont{W.~M.} \bibnamefont{Durham}},
  \bibinfo{author}{\bibfnamefont{E.}~\bibnamefont{Climent}}, \bibnamefont{and}
  \bibinfo{author}{\bibfnamefont{R.}~\bibnamefont{Stocker}},
  \bibinfo{journal}{Phys. Rev. Lett.} \textbf{\bibinfo{volume}{106}},
  \bibinfo{pages}{238102} (\bibinfo{year}{2011}).

\bibitem[{\citenamefont{Khurana and Ouellette}(2012)}]{Khu12}
\bibinfo{author}{\bibfnamefont{N.}~\bibnamefont{Khurana}} \bibnamefont{and}
  \bibinfo{author}{\bibfnamefont{N.~T.} \bibnamefont{Ouellette}},
  \bibinfo{journal}{Phys. Fluids} \textbf{\bibinfo{volume}{24}},
  \bibinfo{pages}{091902} (\bibinfo{year}{2012}).

\bibitem[{\citenamefont{Zhan et~al.}(2014)\citenamefont{Zhan, Sardina, Lushi,
  and Brandt}}]{Zha14}
\bibinfo{author}{\bibfnamefont{C.}~\bibnamefont{Zhan}},
  \bibinfo{author}{\bibfnamefont{G.}~\bibnamefont{Sardina}},
  \bibinfo{author}{\bibfnamefont{E.}~\bibnamefont{Lushi}}, \bibnamefont{and}
  \bibinfo{author}{\bibfnamefont{L.}~\bibnamefont{Brandt}},
  \bibinfo{journal}{J. Fluid Mech.} \textbf{\bibinfo{volume}{793}},
  \bibinfo{pages}{22} (\bibinfo{year}{2014}).

\bibitem[{\citenamefont{De~Lillo et~al.}(2014)\citenamefont{De~Lillo, Cencini,
  Durham, Barry, Stocker, Climent, and Boffetta}}]{Del14}
\bibinfo{author}{\bibfnamefont{F.}~\bibnamefont{De~Lillo}},
  \bibinfo{author}{\bibfnamefont{M.}~\bibnamefont{Cencini}},
  \bibinfo{author}{\bibfnamefont{W.~M.} \bibnamefont{Durham}},
  \bibinfo{author}{\bibfnamefont{M.}~\bibnamefont{Barry}},
  \bibinfo{author}{\bibfnamefont{R.}~\bibnamefont{Stocker}},
  \bibinfo{author}{\bibfnamefont{E.}~\bibnamefont{Climent}}, \bibnamefont{and}
  \bibinfo{author}{\bibfnamefont{G.}~\bibnamefont{Boffetta}},
  \bibinfo{journal}{Phys. Rev. Lett.} \textbf{\bibinfo{volume}{112}},
  \bibinfo{pages}{044502} (\bibinfo{year}{2014}).

\bibitem[{\citenamefont{Gustavsson and Mehlig}(2014)}]{Gus14f}
\bibinfo{author}{\bibfnamefont{K.}~\bibnamefont{Gustavsson}} \bibnamefont{and}
  \bibinfo{author}{\bibfnamefont{B.}~\bibnamefont{Mehlig}},
  \bibinfo{journal}{arxiv:1412.4374}  (\bibinfo{year}{2014}).

\bibitem[{\citenamefont{P\'ecseli et~al.}(2012)\citenamefont{P\'ecseli,
  Trulsen, and Fiksen}}]{Pec12}
\bibinfo{author}{\bibfnamefont{H.}~\bibnamefont{P\'ecseli}},
  \bibinfo{author}{\bibfnamefont{J.}~\bibnamefont{Trulsen}}, \bibnamefont{and}
  \bibinfo{author}{\bibfnamefont{{\O}.}~\bibnamefont{Fiksen}},
  \bibinfo{journal}{Progress in Oceanography} \textbf{\bibinfo{volume}{101}},
  \bibinfo{pages}{14} (\bibinfo{year}{2012}).

\bibitem[{\citenamefont{Schumacher et~al.}(2014)\citenamefont{Schumacher,
  Scheel, Krasnov, Donzis, Yakhot, and Sreenivasan}}]{Sch14}
\bibinfo{author}{\bibfnamefont{J.}~\bibnamefont{Schumacher}},
  \bibinfo{author}{\bibfnamefont{J.}~\bibnamefont{Scheel}},
  \bibinfo{author}{\bibfnamefont{D.}~\bibnamefont{Krasnov}},
  \bibinfo{author}{\bibfnamefont{D.}~\bibnamefont{Donzis}},
  \bibinfo{author}{\bibfnamefont{V.}~\bibnamefont{Yalhot}}, \bibnamefont{and}
  \bibinfo{author}{\bibfnamefont{K.}~\bibnamefont{Sreenivasan}},
  \bibinfo{journal}{PNAS} \textbf{\bibinfo{volume}{11}}, \bibinfo{pages}{10961}
  (\bibinfo{year}{2014}).


\bibitem[{\citenamefont{Yamazaki and Squires}(1996)}]{Yam96}
\bibinfo{author}{\bibfnamefont{H.}~\bibnamefont{Yamazaki}} \bibnamefont{and}
  \bibinfo{author}{\bibfnamefont{K.~D.} \bibnamefont{Squires}},
  \bibinfo{journal}{Mar. Ecol. Prog. Series} \textbf{\bibinfo{volume}{144}},
  \bibinfo{pages}{299} (\bibinfo{year}{1996}).

\bibitem[{\citenamefont{Gargett}(1989)}]{Gar89}
\bibinfo{author}{\bibfnamefont{A.E.}~\bibnamefont{Gargett}},
  \bibinfo{journal}{Ann. Rev. Fluid Mech.} \textbf{\bibinfo{volume}{21}},
  \bibinfo{pages}{419} (\bibinfo{year}{1989}).

\bibitem[{\citenamefont{Gustavsson and Mehlig}(2011)}]{Gus11a}
\bibinfo{author}{\bibfnamefont{K.}~\bibnamefont{Gustavsson}} \bibnamefont{and}
  \bibinfo{author}{\bibfnamefont{B.}~\bibnamefont{Mehlig}},
  \bibinfo{journal}{Europhys.~Lett.} \textbf{\bibinfo{volume}{96}},
  \bibinfo{pages}{60012} (\bibinfo{year}{2011}).


\bibitem[{\citenamefont{Gustavsson
  et~al.}(2014{\natexlab{b}})\citenamefont{Gustavsson, Vajedi, and
  Mehlig}}]{Gus14e}
\bibinfo{author}{\bibfnamefont{K.}~\bibnamefont{Gustavsson}},
  \bibinfo{author}{\bibfnamefont{S.}~\bibnamefont{Vajedi}}, \bibnamefont{and}
  \bibinfo{author}{\bibfnamefont{B.}~\bibnamefont{Mehlig}},
  \bibinfo{journal}{Phys. Rev. Lett.} \textbf{\bibinfo{volume}{112}},
 \bibinfo{pages}{214501}  (\bibinfo{year}{2014}{\natexlab{b}}).


\bibitem[{\citenamefont{Gustavsson et~al.}(2014)\citenamefont{Gustavsson,
  Berglund, Jonsson, and Mehlig}}]{supp}
  \bibinfo{title}{See Supplemental Material at [http://link.aps.org/ 405
supplemental/10.1103/PhysRevLett.000.000000] for details of the calculations, and for a Supplemental Figure.}).

\obs{\bibitem[{\citenamefont{Fouxon et~al.}(2015)\citenamefont{Fouxon,
  and Leshansky}}]{Fou15}
\bibinfo{author}{\bibfnamefont{I.}~\bibnamefont{Fouxon}},
  \bibinfo{author}{\bibfnamefont{A. M.}~\bibnamefont{Leshansky}},
  \bibinfo{journal}{Phys.~Rev.~E} \textbf{\bibinfo{volume}{92}},
  \bibinfo{pages}{013017} (\bibinfo{year}{2015}).}

\bibitem[{\citenamefont{Andersson et~al.}(2007)\citenamefont{Andersson,
  Gustavsson, Mehlig, and Wilkinson}}]{And07}
\bibinfo{author}{\bibfnamefont{B.}~\bibnamefont{Andersson}},
  \bibinfo{author}{\bibfnamefont{K.}~\bibnamefont{Gustavsson}},
  \bibinfo{author}{\bibfnamefont{B.}~\bibnamefont{Mehlig}}, \bibnamefont{and}
  \bibinfo{author}{\bibfnamefont{M.}~\bibnamefont{Wilkinson}},
  \bibinfo{journal}{Europhys.~Lett.} \textbf{\bibinfo{volume}{80}},
  \bibinfo{pages}{69001} (\bibinfo{year}{2007}).

\bibitem[{\citenamefont{Kaplan and Yorke}(1979)}]{Kap79}
\bibinfo{author}{\bibfnamefont{J.}~\bibnamefont{Kaplan}} \bibnamefont{and}
  \bibinfo{author}{\bibfnamefont{J.~A.} \bibnamefont{Yorke}},
  \bibinfo{journal}{Springer Lecture Notes in Mathematics}
  \textbf{\bibinfo{volume}{730}}, \bibinfo{pages}{204} (\bibinfo{year}{1979}).

\bibitem[{\citenamefont{Wilkinson and Mehlig}(2005)}]{Wil05}
\bibinfo{author}{\bibfnamefont{M.}~\bibnamefont{Wilkinson}} \bibnamefont{and}
  \bibinfo{author}{\bibfnamefont{B.}~\bibnamefont{Mehlig}},
  \bibinfo{journal}{Europhys.~Lett.} \textbf{\bibinfo{volume}{71}},
  \bibinfo{pages}{186} (\bibinfo{year}{2005}).

\bibitem[{\citenamefont{Falkovich et~al.}(2002)\citenamefont{Falkovich, Fouxon,
  and Stepanov}}]{Fal02}
\bibinfo{author}{\bibfnamefont{G.}~\bibnamefont{Falkovich}},
  \bibinfo{author}{\bibfnamefont{A.}~\bibnamefont{Fouxon}}, \bibnamefont{and}
  \bibinfo{author}{\bibfnamefont{G.}~\bibnamefont{Stepanov}},
  \bibinfo{journal}{Nature} \textbf{\bibinfo{volume}{419}},
  \bibinfo{pages}{151} (\bibinfo{year}{2002}).

\bibitem[{\citenamefont{Gustavsson et~al.}(2012)\citenamefont{Gustavsson,
  Meneguz, Reeks, and Mehlig}}]{Gus12}
\bibinfo{author}{\bibfnamefont{K.}~\bibnamefont{Gustavsson}},
  \bibinfo{author}{\bibfnamefont{E.}~\bibnamefont{Meneguz}},
  \bibinfo{author}{\bibfnamefont{M.}~\bibnamefont{Reeks}}, \bibnamefont{and}
  \bibinfo{author}{\bibfnamefont{B.}~\bibnamefont{Mehlig}},
  \bibinfo{journal}{New~J.~Phys.} \textbf{\bibinfo{volume}{14}},
  \bibinfo{pages}{115017} (\bibinfo{year}{2012}).

\bibitem[{\citenamefont{Gustavsson and Mehlig}(2004)}]{Meh04}
\bibinfo{author}{\bibfnamefont{K.}~\bibnamefont{Gustavsson}} \bibnamefont{and}
  \bibinfo{author}{\bibfnamefont{B.}~\bibnamefont{Mehlig}},
  \bibinfo{journal}{Phys.~Rev.~E} \textbf{\bibinfo{volume}{87}},
  \bibinfo{pages}{023016} (\bibinfo{year}{2013}).

\bibitem[{\citenamefont{Rothschild and Osborn}(1988)}]{Rot88}
\bibinfo{author}{\bibfnamefont{B.~J.} \bibnamefont{Rothschild}}
  \bibnamefont{and} \bibinfo{author}{\bibfnamefont{T.~R.}
  \bibnamefont{Osborn}}, \bibinfo{journal}{J. Plankton Res.}
  \textbf{\bibinfo{volume}{10}}, \bibinfo{pages}{465} (\bibinfo{year}{1988}).


\end{thebibliography}
\end{document}


\title{Supplemental material for \lq A model for preferential sampling and small-scale clustering of gyrotactic micro-swimmers in turbulence\rq{}}
\author{K. Gustavsson$^{1,2)}$, F. Berglund$^{1)}$, P.R. Jonsson$^{3)}$,
and B. Mehlig$^{1)}$ }
\affiliation{\mbox{}$^{1)}$Department of Physics, Gothenburg University, SE-41296 Gothenburg, Sweden}
\affiliation{\mbox{}$^{2)}$Department of Physics and INFN, University of Rome 'Tor Vergata', 00133 Rome, Italy}
\affiliation{\mbox{}$^{3)}$Department of Biological and Environmental Sciences - Tj\"arn\"o, SE-45296 Str\"omstad,Sweden}

\maketitle

In this Supplemental Material we outline the steps that are needed to derive Eqs.~(5,6) in the Letter \cite{Gus16}. This Supplemental Material also describes how we computed an analytical expression for the Lyapunov dimension shown in Fig.~2{\bf a} of the Letter, and it contains a Supplemental Figure (Fig. S1).

\section{Preferential sampling}
The results described in the Letter \cite{Gus16} rest on the perturbation method described in Refs.~\cite{Gus11a,Gus14e,Gus14f}.
In this Section we summarise how we have applied this method to the equations of motion (1,2) in the Letter, in order to compute the average
of the fluid-velocity field and its gradients along particle paths. These averages determine to which extent the particle preferentially sample
certain fluid-velocity configurations.

In the dimensionless variables $t'=t/\tau, \ve r'= \ve r/\eta, \ve u'= \ve u/u_0, \ve\omega'=\ve\omega\tau$ the equations of motion read:
\begin{align}
\eqnlab{eqm_r}
\frac{{\rm d}{\ve r}'}{{\rm d}t'}&=\ku\ve u'+\Phi\ve n\\
\eqnlab{eqm_n}
\frac{{\rm d}n}{{\rm d}t'}&=\ve\omega'\wedge\ve n\\
\ve \omega'&=-(\ve n \wedge \hat {\ve g})/(2\Psi)+\ku\ve \Omega'(\ve r'(t'), t')
+ \ku\Lambda \ve n \wedge  \big[\ma S'(\ve r'(t'), t')\ve n\big]\,.
\end{align}
The dimensionless parameters $\ku=u_0\tau/\eta$, $\Psi=\mathscr{B}/\tau$, and $\Phi=v_{\rm s}\tau/\eta$ are chosen so that all
terms containing the fluid velocity $\ve u$ or its gradients are proportional to $\ku$. This implies as we shall see that a perturbation expansion
about the deterministic ($\ve u=0$) solution  is equivalent to a perturbation expansion in the Kubo number $\ku$.

To expand Eqs.~\eqnref{eqm_r} and \eqnref{eqm_n} in the Kubo number we first require a $\ku$-expansion of the orientation vector $\ve n$:
\begin{align}
\ve n({t'})=\sum_{q=0}^\infty\ve n_{q}({t'})\ku^q\,,
\end{align}
where $\ve n_{q}({t'})$ are time-dependent expansion coefficients. Inserting this ansatz into
 \Eqnref{eqm_n} and identifying terms of order $\ku^q$ we find that the expansion coefficients must satisfy:
\begin{align}
\frac{{\rm d}}{{\rm d}t'} {\ve n}_{q}({t'})&=\frac{1}{2\Psi}\Big(\sum_{j=0}^q\big(\ve n_{q-j}({t'})\cdot\gghat\big)\ve n_{j}({t'})-\delta_{q,0}\gghat\Big)
+\ma B'(\ve r'(t'), t')\ve n_{q-1}({t'})
\nn\\&
-\sum_{j=0}^{q-1}\sum_{k=0}^{q-j-1}\big(\ve n_{j}(t')\cdot\ma B'(\ve r'(t'), t')\ve n_{k}({t'})\big)\ve n_{q-j-k-1}({t'})\,.
\eqnlab{eqm_nKu}
\end{align}
Here $\ma B' = \ma O'+ \Lambda \ma S'$, and $\ma S'$ and $\ma O'$ are the symmetric and antisymmetric parts of the matrix $\ma A'$ of fluid-velocity gradients.
Eq.~(\ref{eq:eqm_nKu}) can be  solved for $\ve n_{q}$ in terms of integrals over solutions $\ve n_{p}$ with $p<q$.
The lowest-order solution in $\ku$, $\ve n_{0}({t'})$, is deterministic. It does not depend on $\ma B$, and is thus independent of the stochastic
fluid-velocity gradients. This deterministic solution $\ve n_{0}(t')$ depends on the initial orientation in a complicated way. But in the limit of large times the dependence on the initial orientation must drop out (provided that $\Psi\ne 0$). Using this fact
we obtain $\ve n_{0}({t'\to\infty})=-\gghat$. We have verified explicitly that the dependence on the initial orientation
drops out for all averages evaluated in the Letter, by
asymptotic expansion of $\ve n_{0}({t'})$ for large values of $t$ and by numerical evaluation of the exact solution for $\ve n_{0}({t'})$.

Using $\ve n_{0}({t')}=-\gghat$ we find the following solution for
the expansion coefficients $\ve n_{q}({t'})$:
\begin{align}
{\ve n}_{q}({t'})&=-\delta_{q,0}\gghat+\int_0^{t'}{\rm d}t_1'\Big[
{\rm e}^{(t_1'-t')/(2\Psi)}\ve f(t_1') + ({\rm e}^{(t_1'-t')/\Psi}-{\rm e}^{(t_1'-t')/(2\Psi)})(\ve f(t_1')\cdot\gghat)\gghat
\Big]
\,,
\eqnlab{sol_nKu}
\end{align}
with
\begin{align}
\label{eq:deff}
\ve f(t')=\frac{1}{2\Psi}\sum_{j=1}^{q-1}\big(\ve n_{q-j}({t'})\cdot\gghat\big)\ve n_{j}({t'})
+\ma B'(\ve r'(t'), t')\ve n_{q-1}({t'})-\sum_{j=0}^{q-1}\sum_{k=0}^{q-j-1}\big({\ve n_{j}({t'})}\cdot\ma B'(\ve r'(t'), t')\ve n_{k}({t'}))\ve n_{q-j-k-1}({t'})\,.
\end{align}
Now we insert the resulting solution for $\ve n(t')$ into the equation of motion (\ref{eq:eqm_r}). This yields:
\begin{align}
\frac{{\rm d}{\ve r}'}{{\rm d}t'} &=\ku\ve u'\big(\ve r'({t'}),t'\big)+\Phi\sum_{q=0}^\infty\ve n_{q}({t'})\ku^q\,.
\eqnlab{eqm_rnKu}
\end{align}
This equation has the implicit solution
\begin{align}
\ve r'(t')&=\ve r'_{\!\rm det}(t')+\ku\int_0^{t'}{\rm d}t_1'[\ve u'\big(\ve r'(t_1'),t_1'\big)+\Phi\sum_{q=1}^\infty\ve n_{q}({t_1'})\ku^{q-1}]\,.
\eqnlab{rsol_implicit}
\end{align}
Here $\ve r'_{\!\rm det}({t'})$ denotes the deterministic solution, obtained by letting $\ku=0$ in \Eqnref{eqm_rnKu} and using $\ve n_0(t')= -\gghat$:
\begin{align}
\ve r'_{\!\rm det}({t'})=\ve r'(0)-\ku\Phi\gghat t'\,.
\end{align}
Now consider an actual particle path $\ve r'(t')$.
 The difference $\delta\ve r'({t'})\equiv\ve r'({t'})-\ve r'_{\!\rm det}({t'})$ between this path and its deterministic approximation
is obtained from \Eqnref{rsol_implicit}. We see that $\delta\ve r'(t')\sim\ku$. This ensures
that $\delta\ve r'(t')$ is small if $\ku$ is small enough, and it allows us to expand
the fluid velocity $\ve u'\big(\ve r'({t'}),t'\big)$ in $\delta\ve r'$ around the deterministic trajectory $\ve r'_{\!\rm det}({t'})$:
\begin{align}
\ve u'\big(\ve r'({t'}),t'\big)=\ve u'\big(\ve r'_{\!\rm det}({t'}),t'\big)+\big[\delta\ve r'({t'})\cdot\ve\nabla'\big]\ve u'\big(\ve r'_{\!\rm det}({t'}),t'\big)+\dots\,.
\eqnlab{u_expansion}
\end{align}
Inserting this expansion and the corresponding expansion for $\ma B'\big(\ve r'(t'),t'\big)$ in (\ref{eq:deff}) into \Eqnref{rsol_implicit}
gives an expression for $\delta\ve r'(t')$ in terms of $\ve u'\big(\ve r'_{\!\rm det}({t'}),t'\big)$ and its derivatives, analogous to the expressions discussed
for the inertial-particle problem in Ref.~\cite{Gus14f}.

Finally, to evaluate the average $\langle\ve u\rangle_\infty$ we iteratively substitute $\delta\ve r'(t')$
into \Eqnref{u_expansion}, and cut off the resulting series at the desired order in $\ku$.  To first order in $\ku$ we find for example:
\begin{align}
\ve u'\big(\ve r'({t'}),t'\big)&=\ve u'\big(\ve r'_{\!\rm det}({t'}),t'\big)
+\ku\int_0^{t'}{\rm d}t_1'\Big\{\big[\ve u'\big(\ve r'({t_1'}),t_1'\big)+\Phi\ve n_1({t_1'})\big]\cdot\ve\nabla'\Big\}\ve u'\big(\ve r'_{\!\rm det}({t'}),t'\big)\,,
\eqnlab{u_expanded}
\end{align}
with
\begin{align}
{\ve n}_1{(t_1')}&=\int_0^{t_1'}{\rm d}t_2' {\rm e}^{(t_2'-t_1')/(2\Psi)}
\Big\{-\ma B'\big(\ve r'({t_2'}),t_2'\big)\gghat+\big[\gghat\cdot\ma B'\big(\ve r'({t_2'}),t_2'\big)\gghat\big]\gghat\Big\}
\,.
\label{eq:nexp}
\end{align}
Up to this point the expressions are valid for fluid-velocity fields with arbitrary distribution. Now we make use of the fact
that the fluid-velocity field is assumed to be Gaussian in the statistical
model with, zero mean, Gaussian spatial correlations, and with exponential time
correlations \cite{Gus14f}:
\begin{align}
\langle u'_i(\ve r'_1,t'_1)u'_j(\ve r'_2,t'_2)\rangle=
\frac{1}{d(d-1)}\left[\delta_{ij}(d-1 - (\ve r'_1-\ve r'_2)^2) +(\ve r'_1-\ve r'_2)_i(\ve r'_1-\ve r'_2)_j\right]{\rm e}^{-(\sve r'_1-\sve r'_2)^2/2-|t_1'-t_2'|}\,.
\eqnlab{uCorrFun}
\end{align}
The form of the prefactor follows from isotropy, homogeneity, incompressibility, and normalisation $\langle\ve u'(\ve 0,0)^2\rangle=1$.
Taking the average of Eqs.~(\ref{eq:u_expanded}) and
(\ref{eq:nexp}) we find:
\begin{align}
&\langle\ve u'\rangle_\infty=-\gghat\frac{\ku\G}{d}\Big\{\frac{C_0}{2 \G^2}
+\frac{\Psi C_{\G}}{\sqrt{2}\G^3}\E\Big[\frac{1}{\sqrt{2} \G}\Big]
-\frac{4 \Psi^2 C_{\G}+\left(4\Psi+1\right)C_0}{4\sqrt{2}\G^3 \Psi}\E\Big[\frac{2\Psi+1}{\sqrt{8}\G\Psi}\Big] \Big\}\,,
\eqnlab{u_average}
\end{align}
with $C_{x}\equiv\Lambda+1+(d+1)(\Lambda-1)x^2$ and $\E[x]\equiv\sqrt{\pi}e^{x^2}\erfc[x]$.
In the same way we obtain the average of the fluid-velocity gradient matrix $\ma A'$:
\begin{align}
&\langle \ma A'\rangle_\infty=\frac{\ku\G}{d}
\frac{\delta_{ij}-d\hat g_i\hat g_j}{1-d}
\Big\{
C_0\frac{4\Psi+1}{4 \G^3 \Psi}
+\frac{\Psi C_{\G}}{\sqrt{2}\G^4}\E\Big[\frac{1}{\sqrt{2} \G}\Big]
-\frac{(2\Psi+1)\left[4 \Psi^2 C_{\G}+\left(4\Psi+1\right)C_0\right]}{8\sqrt{2} \G^4 \Psi^2}\E\Big[\frac{2\Psi+1}{\sqrt{8}\G\Psi}\Big]
\Big\}\,.
\eqnlab{A_average}
\end{align}
The $z$-components of Eqs.~\eqnref{u_average} and \eqnref{A_average} are plotted in Fig.~1 in the Letter \cite{Gus16}.
The asymptotic behaviour for small values of $\G$ is obtained by series expansion in $\Phi$:
\begin{align}
\langle A'_{zz}\rangle_\infty&\sim \ku\G^2\frac{d(1-\Lambda) + 2(\Lambda+2)}{d}\,\frac{\Psi(4\Psi+1)}{(2\Psi+1)^2}\,,\\
\langle u_z'\rangle_\infty&\sim -\,\ku\G\frac{d(1-\Lambda)+2}{d}\,\frac{\Psi}{2\Psi+1}\,.
\end{align}
These equations correspond to Eqs.~(5a) and (5b) in the Letter.

A corresponding series expansion for large $\G$ of Eqs.~\eqnref{u_average} and \eqnref{A_average} gives
\begin{align}
\eqnlab{S_largeG}
\langle A'_{zz}\rangle_\infty&\sim \frac{\ku}{\G}\frac{d+1}{2d}(1-\Lambda)\sqrt{\frac{\pi}{2}}\,,\\
\eqnlab{uz_largeG}
\langle u_z'\rangle_\infty&\sim \frac{\ku}{\G} \frac{(d(\Lambda-1) + 2\Lambda)}{2d}\,.
\end{align}
These are Eqs.~(6a) and (6b) in the Letter.  We note that this result
can also be found by an expansion for large $\G$ of the original equations for an arbitrary homogeneous, isotropic and incompressible flow, for arbitrary values of $\ku$. Because this limit is universal it is not necessary to assume that $\ku$ is small.

\newpage\section{Small-scale clustering}
The small-scale spatial patterns of gyrotactic micro-swimmers are fractal \cite{Dur13}.
Fractal small-scale clustering is quantified by fractal dimensions (see Ref.~\cite{Gus14f} for a review). There are different
fractal dimensions, characterising different geometrical aspects of the spatial distribution. In the Letter we show results for the fractal Lyapunov dimension in $d=2$ spatial dimensions. Eq.~(11) in the Letter shows that this dimension is defined in terms of the Lyapunov exponents:
\begin{align}
\lambda_1\tau&\equiv\lim_{t'\to\infty}{\frac{1}{t'}}\ln \frac{R'({t'})}{R'(0)}
\quad\mbox{and}\quad (\lambda_1+\lambda_2)\tau\equiv\lim_{t'\to\infty}{\frac{1}{t'}}\ln \frac{{\cal A'}({t'})}{{\cal A'}(0)}\,.
\end{align}
The exponent $\lambda_1$ gives the rate which the separation $R'(t')$ between two initially nearby   micro-swimmers decreases or increases exponentially at long times.  The sum $\lambda_1+\lambda_2$ determines the rate at which the infinitesimal area $A'(t')$ of the parallelogram spanned by the separation vectors between
three initially nearby   micro-swimmers decreases or increases at long times. The problem of computing small-scale fractal clustering
in $d$ dimensions is a $d+1$-particle problem.

We solve this problem using the method described in Ref.~\cite{Gus14f}.
Linearising the equation of motion \Eqnref{eqm_r} we see that the
separation between two initially nearby   micro-swimmers evolves according to
\begin{align}
\label{eq:Rdot}
\frac{{\rm d} R'}{{\rm d}t'} &=\ku\,(\hat{\ve R} \cdot \ma Z'\hat{\ve R})\,R'\,,
\end{align}
where $\hat{\ve R}$ denotes the unit vector along the separation direction
between the two micro-swimmers, and $\ma Z'=\ma A'+\G\ma Y'$ is the particle-velocity gradient matrix. Here $\ma Y'$ is the matrix of orientation gradients with elements $Y_{ij}'= \partial n_i/\partial r'_j$.
It follows from Eq.~(\ref{eq:Rdot}) that the exponent $\lambda_1$ is given by
\begin{align}
\lambda_1\tau&=\ku\,\langle \hat{\ve R}\cdot \ma Z'\hat{\ve R}\rangle_\infty\,.
\end{align}
In two spatial dimensions the sum of $\lambda_1$ and $\lambda_2$ is simply
\begin{align}
(\lambda_1+\lambda_2)\tau=\ku\,\langle\tr\ma Z'\rangle_\infty\,.
\end{align}
To evaluate these averages we require the time evolution of
$\hat{\ve R}$ and $\ma Y'$.
This follows from linearisation of Eqs.~\eqnref{eqm_r} and \eqnref{eqm_n}:
\begin{align}
\eqnlab{eqm_dr}
\frac{{\rm d}\hat{R}_i}{{\rm d}t'}
&=\ku[Z'_{ij} \hat{R}_j-(\hat{R}_jZ'_{jk}\hat{R}_k)\hat{R}_i]\,,\\
\frac{{\rm d} Y'_{ij}}{{\rm d}t'} &=
\frac{1}{2\Psi}[\ghat_k Y'_{kj}n_i+n_k\ghat_kY'_{ij}]
+\ku[(\partial_jB'_{ik})n_k+B'_{ik}Y'_{kj}
-2\Lambda(n_kS'_{kl}Y'_{lj})n_i
-\Lambda(n_k(\partial_jS'_{kl})n_l)n_i
\nn\\&
-\Lambda(n_kS'_{kl}n_l)Y'_{ij}
-Y'_{ik}(A'_{kj}+\Phi Y'_{kj})
]\,,
\eqnlab{eqmY}
\end{align}
where repeated indices are summed over. We solve \Eqnref{eqmY} in terms of a series expansion
\begin{align}
\ma Y'({t'})=\sum_{q=0}^\infty\ma Y'_{q}({t'})\ku^q\,.
\end{align}
Inserting this expression into \Eqnref{eqmY}
we find the following expression for the expansion coefficients:
\begin{align}
Y_{q;ij}(t')&=
\int_0^{t'}{\rm d}t_1'\left[
{\rm e}^{(t_1'-t')/(2\Psi)}F_{ij}(t_1')
+({\rm e}^{(t_1'-t')/\Psi}-{\rm e}^{(t_1'-t')/(2\Psi)})\ghat_i\ghat_k F_{kj}(t_1')
\right]\,,
\end{align}
where we the initial condition was put to zero because the corresponding terms do not contribute to steady-state averages.
The coefficients $F_{ij}(t')$ are given by
\begin{align}
F_{ij}(t')&=
\frac{1}{2\Psi}\sum_{r=0}^{q-1}[\ghat_kY_{r;kj}' n_{q-r;i}+ n_{q-r;k}\ghat_k Y_{r;ij}' +\delta_{q,1}[(\partial_jB'_{ik}(\ve r'(t'),t'))n_k-\Lambda n_k(\partial_jS'_{kl}(\ve r'(t'),t')n_l)n_i
\nn\\&
+B'_{ik}(\ve r'(t'),t') Y'_{kj}
-2\Lambda n_kS'_{kl}(\ve r'(t'),t') Y'_{lj}n_i
-\Lambda n_kS'_{kl}(\ve r'(t'),t')n_l Y'_{ij}
-Y'_{ik}A'_{kj}(\ve r'(t'),t')
-\Phi Y'_{ik}Y'_{kj}
]\,.
\eqnlab{Y_implicit}
\end{align}
\Eqnref{eqm_dr} can be solved implicitly as
\begin{align}
\hat{R}_i(t')&=\ku\int_0^{t'}{\rm d}t_1'\Big\{Z'_{ij}(t_1') \hat{R}_j(t_1')-
\big[\hat{R}_j(t_1')Z'_{jk}(t_1')\hat{R}_k(t_1')\big]\hat{R}_i(t_1')\Big\}\,.
\eqnlab{dr_implicit}
\end{align}
By substituting $\hat{\ve R}$ and $\ma Y'$ in the right-hand side of Eqs.~\eqnref{Y_implicit} and \eqnref{dr_implicit}, Eqs.~(\ref{eq:Y_implicit}) and (\ref{eq:dr_implicit})  can be recursively iterated to any desired order in $\ku$.
Expanding the fluid velocity and gradients of the fluid velocity as in the previous Section gives equations for $\hat{\ve R}$ and $\ma Y'$ in terms of the fluid velocity evaluated along the deterministic trajectories.
Averaging gives the Lyapunov exponents to order $\ku^2$
 \begin{align}
\eqnlab{lambda1Res}
&\lambda_1\tau=\frac{{\rm Ku}^2}{d(d-1)\G^5}\Big\{\Big[(d\!-\!3)\G^2\! +\! (3\!-\!d\!+\!(15\!-\!4d\!+\!d^2)\G^2)\langle(\hat{\ve R}\cdot\hat{\ve g})^2\rangle\!-\!2(1\!+\!5\G^2)\langle(\hat{\ve R}\cdot\hat{\ve g})^4\rangle\Big]\G\! +\!\frac{1}{\sqrt{2}}\Big[(3\!-\!d)\G^2 \\&
+(3\!-\!2d\!+\!d^2)\G^4\!-\!(3\!-\!d\!+\!(18\!-\!5d\!-\!d^2)\G^2\!+\!(9\!-\!2d\!+\!d^2)\G^4)\langle(\hat{\ve R}\cdot\hat{\ve g})^2\rangle
+ (2+12\G^2+6\G^4)\langle(\hat{\ve R}\cdot\hat{\ve g})^4\rangle \Big] \E[{1}/{(\sqrt{2}\G)}] \Big\}\nn\,,
\\
&(\lambda_1+\lambda_2)\tau=
 \frac{\ku^2\Psi^2}{d\G^3}\big\{
 (\Lambda+1)\left[C_{\G}+\G^2 (d (\Lambda-1)-2)\right]\G
-\big[C_{\G}^2+2(d+1)(\Lambda-1)^2\G^4\big]\E\big[{1}/{(\sqrt{2} \G)}\big]/\sqrt{2}
 \big\}\,.
 \eqnlab{TrZRes}
 \end{align}
The steady-state moments $\langle(\hat{\ve R}\cdot\hat{\ve g})^{2p}\rangle_\infty$ in \Eqnref{lambda1Res}
characterise the anisotropy of the spatial patterns.
The moments are given by Eq.~(8) derived in Ref.~\cite{Gus14e} for particle pairs settling in a turbulent aerosol,  upon replacing ${\rm G}$ in that equation by $-\G$. The fact that the moments agree is a
coincidence, only valid to lowest order in $\ku$.  While the problems are superficially related (to lowest
order particles move on a deterministic trajectory $\ve r'_{\!\rm det}(t')$ through turbulence) the details are quite different.
Using the above expressions for the Lyapunov exponents (with $\lambda_1$ extended to order $\ku^4$) we obtain the curves plotted in Fig. 2{\bf a} in the Letter.

Series expansions of \Eqnref{TrZRes} for small and large values of $\Phi$ give Eqs.~(7) and~(9) in the Letter
\begin{align}
\langle \ve \nabla \cdot \ve v\rangle_\infty\eta/u_0&\sim-\ku\, (\Phi\Psi)^2 B_d(\Lambda)\,\;\mbox{for}\;\Phi\ll 1\,,\\
\langle \ve \nabla \cdot \ve v\rangle_\infty\eta/u_0&\sim-\ku\, \Phi\Psi^2 E_d(\Lambda)\,\;\mbox{for}\;\Phi\gg 1\,.
\end{align}
The shape-dependent prefactors $B_d(\Lambda)$ and $E_d(\Lambda)$ are found to be:
\begin{align}
B_d(\Lambda)&\equiv [(d+2)(d+4) - 2d(d+4)\Lambda+ (4+2d+d^2)\Lambda^2\big]/d\\
E_d(\Lambda)&\equiv\sqrt{\pi/2}(d+1)(d+3)(\Lambda-1)^2/d\,.
\end{align}
The corresponding limits for $\lambda_1$ are more subtle due to the complicated dependence of the moments $\langle(\hat{\ve R}\cdot\hat{\ve g})^{2p}\rangle_\infty$ on $\Phi$.
For small values of $\Phi$ we have $\langle(\hat{\ve R}\cdot\hat{\ve g})^{2p}\rangle_\infty\sim(2p-1)!!/(2^pp!)$~\cite{Gus14e} and consequently
\begin{align}
\lambda_1\tau\sim -2\frac{d^2+d-3}{d(d-1)}\ku^2\Phi^2\,.
\end{align}
This equation is quoted in the Letter for $d=2$.

The results derived here are obtained for general values of $\Phi$, $\Psi$, and $\Lambda$. It is important to note that the limit of large $\Psi$ is singular. In this limit the dependence on the initial orientation $\ve n(t=0)$ does not decay fast enough for it to be disregarded. This explains, as pointed out in the Letter, that the theory must fail for very large $\Psi$. 

We also note that Eqs.~\eqnref{lambda1Res} and \eqnref{TrZRes} are derived assuming that the elements of $\ma Y'$ are small enough, allowing us to expand \Eqnref{eqmY} in $\ku$. This may fail when singularities occur in the dynamics of $\ma Y'$. Elements of $\ma Y$ may repeatedly diverge to $-\infty$. These singularities are analogous to caustics in inertial-particle dynamics \cite{Gus14f} and correspond to instances where the $\ve n$-field becomes multi-valued so that nearby particles can have very different orientations.  The perturbation expansion leading to  Eqs.~\eqnref{lambda1Res} and \eqnref{TrZRes} is expected to diverge when caustics are frequent, precisely as in the inertial-particle problem (this point is  discussed in detail in Ref.~\cite{Gus14f}).

\newpage\section{Supplemental Figure}
\begin{figure}[h]
  \begin{overpic}[width=8.5cm,clip]{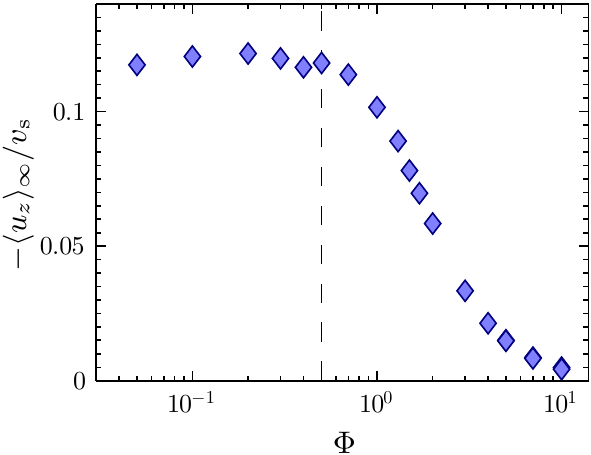}
  \end{overpic}\hspace*{2mm}
  \caption{\label{fig:plateau}
  Shows $\langle u_z\rangle_\infty$ as a function of $\Phi$ for $\ku=1$, $d=3$, $\Psi=1$, and $\Lambda=0$. The data used in this Figure are identical to the data shown in Fig.~1{\bf d} in the Letter. The vertical dashed line shows the value of $\Phi$ used for the $\ku\!=\!1$-data ($\marker{5}$) in Fig.~1{\bf f} in the Letter.}
\end{figure}

Fig.~\ref{fig:plateau} shows that $\langle u_z\rangle_\infty/v_{\rm s}$ approaches a constant as $\Phi\rightarrow 0$.  The parameters are $\ku=1$, $d=3$, $\Psi=1$, and $\Lambda=0$. The Figure demonstrates that the plateau forms also for larger Kubo numbers, not only for small Kubo numbers (Fig.~1{\bf e} in the Letter shows a corresponding plateau for $\ku=0.1$).

